\documentclass[12pt]{article}
\usepackage{graphics}
\usepackage{graphicx}
\usepackage{color}
\begin{document}
\def\bc{\begin{center}}
\def\ec{\end{center}}
\def\bq{\begin{quotation}}
\def\eq{\end{quotation}}
\def\be{\begin{equation}}
\def\ee{\end{equation}}
\def\bearr{\begin{eqnarray}}
\def\eearr{\end{eqnarray}}
\def\bfl{\begin{flushleft}}
\def\efl{\end{flushleft}}
\def\bfr{\begin{flushright}}
\def\efr{\end{flushright}}
\def\a{\alpha}
\def\b{\beta}
\def\d{\delta}
\def\g{\gamma}
\def\k{\kappa}
\def\w{\omega}
\def\s{\sigma}
\def\Ups{\Upsilon}
\def\del{\partial}
\newcommand{\comment}[1]{}
\bfl
\today \hspace*{3.8in} hep-ph/0611156  
\efl
\bc
{\large\bf
GRAVITON RESONANCES IN {\Large\boldmath $e^+e^- \to \mu^+\mu^-$} AT LINEAR 
COLLIDERS WITH BEAMSTRAHLUNG \& ISR EFFECTS} \\
\vskip 5pt
{\large\sl
Rohini M. Godbole\footnote{Centre for High Energy Physics, Indian 
Institute of Science, Bangalore 560012, India.  \\ \hspace*{0.2in}
Electronic address: {\sf rohini@cts.iisc.ernet.in}}, 
Santosh Kumar Rai\footnote{Harish-Chandra Research Institute, Chhatnag 
Road, Jhunsi,-Allahabad 211019, India. \\ \hspace*{0.2in} Electronic 
address: {\sf skrai@mri.ernet.in}},
{\rm and} Sreerup Raychaudhuri\footnote{Department of Physics, Indian 
Institute of Technology, Kanpur 208016, India. \\ \hspace*{0.2in}
Electronic address: {\sf sreerup@iitk.ac.in}} 
}
\vskip 10pt
{\large\bf Abstract}
\vskip -10pt
\ec
\bq \noindent 
{\small Electromagnetic radiation emitted by the colliding beams is expected to
play an important role at the next generation of high energy $e^+e^-$ linear
collider(s). Focusing on the simplest process ${ e^+e^- \to \mu^+\mu^-}$, we show
that radiative effects like initial state radiation (ISR) and beamstrahlung can
lead to greatly-enhanced signals for resonant graviton modes of the
Randall-Sundrum model. }
\eq 

                     \section{Introduction}

\noindent The next generation of high-energy $e^+ e^-$
colliders~\cite{tesla,others} will necessarily have linear design in order to
avoid crippling energy losses from synchrotron radiation. Obviously any linear
collider will have single-pass colliding beams, unlike a storage ring where
multiple bunch-crossings are possible. High luminosities at linear machines can
only be achieved, therefore, by using beams with bunches of high number density
--- which, in practice, means that the bunches must be focussed to very small
sizes.  Thus, as compared to a bunch length of around 100~$\mu$m at the
now-decommissioned LEP Collider, the bunches at the planned TESLA are expected to
be around $553~{\rm nm} \times 5~{\rm nm} \times 300~{\rm nm} $, while at the CLIC
the planned size is $202~{\rm nm} \times 2.5~{\rm nm} \times 30~{\rm nm}$.  In
fact, narrow, intense beams of this kind constitute a basic and unavoidable part
of the design of all the proposed high-energy $e^+e^-$ machines including the
International Linear Collider (ILC), which is now being planned and designed
through a major international effort~\cite{ILC}.

\bigskip\noindent Although high number densities per bunch do enable the
machine(s) in question to achieve the required luminosity, this feature of the
design is not unaccompanied by its own set of problems. These arise because very
high densities of charged particles at the interaction point will naturally lead
to the generation of strong electromagnetic fields in and around every colliding
bunch. Interaction of beam constituents ($e^\pm$) with the (strong) accelerating
field is well known to generate the so-called {\it initial state radiation} (ISR),
which is essentially a bremsstrahlung phenomenon. At the same time, their
interaction with the strong magnetic field generated {\sl by the other beam} will
also generate radiation, which goes by the name of {\it
beamstrahlung}~\cite{beamold}. In either case, the effect is to make the initial
$e^\pm$ radiate one or multiple high energy photons, which not only tend to
diminish the beam energy, but can cause further complications by behaving like an
extra weak photon beam alongside the electron (positron)  one. Thus, precise
knowledge of the ISR plus beamstrahlung photon spectrum -- and, by inference, of
the spectrum of the emitting electron (positron) after photon emission -- is
essential if we wish to make any realistic predictions at a linear collider.

\bigskip\noindent Fortunately, the spectral distribution of radiated photons can
be accurately calculated using only quantum electrodynamics (QED), which may
require computation to higher orders of perturbation theory (depending on the
desired accuracy), but is eventually extremely precise in its predictions.  In
fact, over the years, starting with the pioneering work of Weiz\"acker and
Williams~\cite{weizwill}, more and more refined calculations of the radiated
photon spectrum have been developed and most of the recent developments are
readily available in the literature\cite{beamstr}.

\bigskip\noindent It has long been a part of the folklore of collider physics to
consider radiation effects like ISR and beamstrahlung to be a nuisance, insofar as
they cause losses in energy and disrupt the beam collimation.  This is, in fact,
what happens when the machine energy is tuned to a resonance, as, for example, was
the case at the LEP-1 collider, which had $\sqrt{s} \simeq M_Z$. At LEP-1,
however, because of the low number density per bunch, beamstrahlung effects were
negligible. At higher energies, where the beamstrahlung effects are not so weak,
radiation-engendered photon beams may, in principle, give rise to a whole new
class of backgrounds due to production of lepton pairs \cite{chen} and hadrons
\cite{dreesRG}. Most machine designs, therefore, try to ensure that these
radiation effects are minimised --- a common trick being to generate {\sl
flattened} bunches, which can be shown, other conditions remaining the same, to
produce minimum radiation\footnote{Such beam designs are certainly true of the
bunch sizes (mentioned above) at the TESLA and the CLIC.}. In this article,
however, we pursue the counter-argument that far from being a nuisance, ISR and
beamstrahlung can sometimes act as a blessing in disguise, since they are useful
in probing new physics scenarios.

\bigskip\noindent Use of radiated photons in new physics studies is by no means a
novel idea. Tagging with large-$p_T$ ISR photons has been used in the LEP
experiments to search for final states which leave (almost) no visible energy in
the detectors -- well-known examples being the neutrino counting processes $e^+e^-
\to \gamma \nu_\ell\bar\nu_\ell$ in the Standard Model, or, in a supersymmetric
model, $e^+e^- \to \gamma\tilde \chi^0 \tilde \chi^0$, where the final state is a
single photon and missing energy\cite{pandita}. The hard photon tag has also been
used in the study of $e^+e^- \to \gamma\tilde \chi^+ \tilde \chi^-$ with
$\tilde\chi^\pm $ and the lightest sparticle $\tilde \chi^0_1$ being almost
degenerate \cite{Chen:1995yu}, as well as searches for gravitonic states similar
to the ones under study in this article\cite{Osland}. For the majority of ISR and
beamstrahlung photons, however, the situation is somewhat different, as they are
mostly collinear with the beams and hence are simply lost down the beam pipe.
Obviously the question of tagging such photons does not arise.  However, since
each photon carries away a different amount of energy from its parent (radiating)
particle, the radiation effects end up in generating an energy distribution of the
initially monochromatic electron and positron beams.  This energy distribution can
be used with profit to search for resonances.

\bigskip\noindent Energy spread of the beams can have a wide variety of
consequences. In order to have a focussed discussion, however, we concentrate on
one of the simplest and cleanest processes at an $e^+ e^-$ collider, viz.
$$ 
e^+ ~e^- \to X^* \to \mu^+ ~\mu^- 
$$ 
where $X$ can be either a massive scalar, vector or tensor particle.  In the
Standard Model (SM), the only options are $X = \gamma, Z$.  For any heavy particle
$X$, there will be resonances in the $s$-channel process, observable as peaks in
the invariant mass $M_{\mu^+\mu^-}$ distribution. At LEP-1, this process, among
others, was used to measure the $Z$-resonance line shape. Heavy particles $X$
predicted in models beyond the Standard Model would show up as similar peaks at a
different value of $M_{\mu^+\mu^-}$. These could be, for example, extra $Z'$
bosons or leptophilic scalars, such as the $\widetilde{\nu}_{\tau}$ in
$R$-parity-violating supersymmetry. In this note, however, we discuss the case
where the $X$ are {\sl tensor} particle resonances, the tensors being the massive
Kaluza-Klein graviton excitations predicted in the well-known brane-world model of
Randall and Sundrum~\cite{RS}\footnote{The Randall-Sundrum model is briefly
reviewed in Section~3.}

\bigskip\noindent The main motivation of the present study is the fact that it is
more or less certain that the next-generation linear collider(s) will be run at a
single, or -- at best -- a limited set of fixed values of the centre-of-mass
energy $\sqrt{s}$. For example, in the TESLA studies it has been assumed that the
collider will be run at two different values of $\sqrt{s}=$~500~GeV and 800~GeV
respectively, the ILC~\cite{ILC} is being planned to run at $\sqrt{s} = 500$~GeV
and 1.0~TeV, while the quoted numbers for the CLIC~\cite{CLIC} are usually
$\sqrt{s} = 1.0$~TeV and 3.0~TeV. However, the massive graviton excitations of the
Randall-Sundrum (RS) model, whose masses are not predicted by the theory, may not
lie very close to these exact centre-of-mass energy values.  In such a case, the
contribution due to exchange of RS gravitons will normally be off-resonance and
hence strongly suppressed.  It may appear, therefore, that a high energy $e^+e^-$
collider will simply miss these new physics effects.  However, a spread in
beam-energy --- such as the one that would be induced by beam radiation --- would
cause a certain number of the events to take place at an effective (lower)
centre-of-mass energy, exciting the resonance(s) and hence providing a significant
enhancement in the cross-section. The expected suppression of this cross-section
due to one extra power of $\alpha$ is more than compensated by the large resonant
cross-section, particularly if the resonances are sharp and narrow. A similar
effect, in fact, was observed in $Z$-resonances at LEP-1.5, running at $\sqrt{s}
=$ 130 and 136 GeV, and dubbed the `return to the $Z$-peak'. Following this idea,
therefore, we investigate a `return to the {\it graviton} peak(s)' in the process
$e^+ ~e^- \to \mu^+ ~\mu^-$. This article completes a preliminary
study~\cite{selves} made earlier by the authors, and complements recent
studies~\cite{AseshCakir} which investigate similar techniques to identify scalar
and vector resonances in other models of new physics.

      \section{Electron Luminosity with ISR and Beamstrahlung}

In the present study, the principal quantity of interest is not the energy of the
emitted photons (which are lost down the beam pipe) but the remanent energy of the
electron or positron after the photon emission. This is because the photon is lost
down the beam pipe, but the effective centre-of-mass energy of the $e^+e^-$ system
(and hence of the final states barring the lost photon(s)) will be determined by
this reduced beam energy. Noting that there may be single or multiple photon
emission due to both ISR and beamstrahlung effects, a convenient way to obtain an
energy spectrum for the electron (positron) is to make use of the {\it structure
function formalism}, which has hitherto been used mainly to predict excess
Standard Model backgrounds\cite{chen,rohini}. In this method -- which is inspired
by and closely follows the standard techniques developed for hadronic interactions
-- the scattering cross section for a given process
$$
e^+(p_1) + e^-(p_2) \to F(x_1 p_1 + x_2 p_2) + (\gamma)
$$ 
where $F$ is any (observable) final state, is given by 
\bearr
\s[e^+e^- \to F (\gamma)]
 = \int dx_1dx_2 ~f_{e/e}(x_1)~f_{e/e}(x_2)~\hat{\s}[e^+e^- \to F] (\hat{s})\ ,
  \label{eq:strucfun}
\eearr
where, as in the case of hadron colliders, $x$ is the momentum fraction of the
electron~(positron) in the initial state while the {\sl electron luminosity}
function $f_{e/e}(x)$ describes the probability of a beam electron of energy $E_b
= \sqrt{s}$ emitting one or more photons of total energy $(1 - x)E_b$, thereby
degrading its own energy to $E_e~=~xE_b$.  It is now a simple matter to show that
for colliding beams the effective centre-of-mass energy will be $\sqrt{\hat{s}}
\simeq \sqrt{x_1 x_2 s}$. The final state, in our study, is $F = \mu^+\mu^-$.

\bigskip\noindent The electron luminosity function $f_{e/e}(x)$, introduced above,
must include both the ISR and beamstrahlung effects. The ISR contribution is
usually calculated in terms of the momentum fraction $x$ using a
Weiz\"acker-Williams approximation, which is perturbatively corrected to take into
account emission of multiple soft photons. The result, to one-loop
order\footnote{which is free of collinear divergences, \'a-la Bloch and
Nordsieck~\cite{bloch}} is given in terms of a normalised distribution
function~\cite{kuraev}
\bearr
f^{\rm ISR}_{e/e}(x) = \frac{\b}{16}
\left[ (8 + 3\b) (1 - x)^{\b/2 - 1} - 4 (1 + x) \right] \ ,
\label{eq:ISR}
\eearr
where the $\b$-parameter is 
\bearr
 \b = \frac{2\a}{\pi}\left(\log \frac{s}{m_e^2} - 1 \right) \ , 
\eearr
and $\a$ denotes the running fine-structure constant evaluated at the beam energy
$E_b$.  At linear collider energies, $\b$ varies more-or-less between 0.13---0.15,
indicating that the first term in Eqn.~(\ref{eq:ISR}) has a negative exponent.
This makes the function $f^{\rm ISR}_{e/e}(x)$ rise steeply as $x \to 1$. Of
course, there is no singularity at $x = 1$ because the overall function is
normalised, but nevertheless, this steep increase ensures that the bulk of the
electron luminosity lies above $x > 0.9$. This immediately tells us that ISR is a
rather weak effect insofar as spreading out the beam energy is concerned. A glance
at the dotted line (marked ISR) in Figure~1 makes this amply clear. Consideration
of ISR effects alone will not, therefore, be of much use in exciting a resonance,
unless the latter happens to lie very close to the machine energy (as was the case
with the $Z$-resonance at LEP-1.5).

\begin{figure}[htb]
\begin{center}
\includegraphics[height=3.0in]{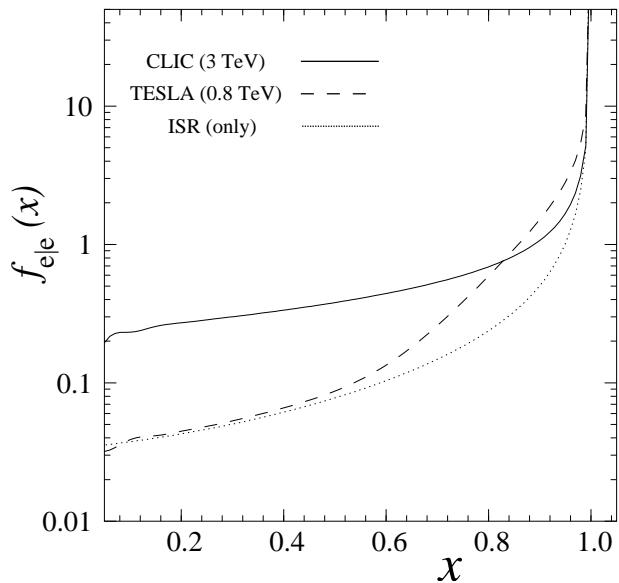}
\end{center}
\vskip -20pt
\caption{\footnotesize \it Illustrating the electron luminosity $f_{e|e}(x)$ as a
function of $x = E_e/E_b$, the energy fraction of the electron (positron) after
radiation of one or more photons. The (dashed) solid line shows the prediction at
the {\em TESLA (CLIC)} machine, where the beamstrahlung parameter is $\Upsilon =
0.09~(8.1)$. The dotted line shows the (unconvoluted) ISR prediction at the {\em
TESLA} energy.}
   \label{fig:bmsisr}
\end{figure}

\bigskip\noindent This situation changes quite spectacularly when we incorporate
beamstrahlung effects. Unlike the ISR spectrum, the beamstrahlung spectrum depends
not only on the electron beam energy $E_b$ and the remanent momentum fraction $x$,
but also on such highly machine-specific parameters as the bunch length $\s_z$ and
the beamstrahlung parameter $\Ups$. The latter, a measure of the effective
magnetic field of the bunches, is given by
\bearr
\Ups = \frac{E_b}{m_e} \left(\frac{B}{B_c}\right) \ ,
\eearr
where $B$ is the effective magnetic field strength in the beam, and $B_c = m_e^2/e
\hbar \simeq 4.4 \times 10^{13}$ Gauss is the Schwinger critical field for
electrons. The magnetic field $B$ will be determined by the density of electrons
(positrons) in the colliding bunches. For beams with a Gaussian energy profile,
the effective or mean value of $\Ups$ is, thus, given by~\cite{chen1}
\bearr
\Ups = \left( \frac{5 r_e^2}{6\a m_e} \right)
~\frac{E_bN_e}{\s_z(\s_x +\s_y)}
\eearr
where $N_e$ is the number of electrons or positrons in a bunch, $\s_x, \s_y$ are
the transverse bunch sizes, and $r_e \simeq 2.818 \times 10^{-13}$~cm is the
classical electron radius. It may be noted that the effective beam sizes $\s_x,
\s_y, \s_z$ which appear in the above equation are slightly different from the
nominal beam sizes (created by focussing magnets) because of the so-called beam
disruption effect, i.e. the tendency of the beams to deform in the strong
electromagnetic fields experienced just before (and during) the collision process.
In terms of these parameters, then, the beamstrahlung spectrum for radiated
photons can be written~\cite{chen1} in terms of the momentum fraction $x$, as
\bearr
f^{\rm beam}_{\g|e}(x) &=&\frac{\k^{1/3}}{\Gamma(\frac{1}{3})}
\frac{e^{-\k x/(1-x)}}{\{x^2 (1-x)\}^{1/3} }  
\left[\frac{1-\w}{\tilde{g}(x)}\left\{1 - 
\frac{1 - e^{-N_\g\tilde{g}(x)}} {\tilde{g}(x)N_\g} \right\}
+ \w\left\{1 - \frac{1 - e^{-N_\g}}{N_\g}\right\}\right],
\eearr
with
\bearr
\tilde{g}(x) = 1 - \frac{1}{2}(1 - x)^{2/3}\left[ 1 - x + (1 + x)
\sqrt{1 + \Ups^{2/3}}\right] \ ,
\qquad \k = \frac{2}{3\Ups} \ , \qquad \w = \frac{1}{6\sqrt{\k}} \ .
\eearr
Here, $\Gamma(\frac{1}{3})$ is the usual Euler gamma function and the quantity
$N_\g$, denoting the average number of photons emitted per electron, is given by
\bearr
N_\g = \left( \frac{5\a^2m_e}{2r_e} \right) 
\left( \frac{\s_z}{E_b} \right)
\frac{\Ups}{\sqrt{1 + \Ups^{2/3}}} \ ,
\eearr

\noindent If we were to consider only single photon emission, the electron
spectrum --- which is what we actually require --- would be just complementary to
the photon spectrum in Eqn.~(6) for $N_\g = 1$. However, we also need to take into
account multiple photon emissions~\cite{chen,chen1,beamstr}. This calculation is
obviously much more involved, but for our purposes it suffices to use a
closed-form approximation for the function $f_{e/e}^{\rm beam} (x)$, viz.,
\bearr
N_\g f^{\rm beam}_{e/e}(x) = (1 - e^{-N_\g})\d(1-x)
+ \frac{e^{-\eta(x)}}{1-x}
\sum_{n=1}^{\infty}\left\{1 - x + x\sqrt{1 + \Ups^{2/3}}\right\}^n 
\frac{\eta(x)^{n/3} \g_{n+1}(N_\g) }{n!~~\Gamma(\frac{n}{3})}
\eearr
where $\eta(x)=\k(\frac{1}{x} -1)$, and $\g_a(s)$ is the incomplete $\g-$function,
defined by
$$
\g_a(x) = \int_0^a dt ~t^{x+1}e^{-t} \ .
$$
The expression in Eqn.~(9) is quite accurate for low values of $\Ups$ up to about
$\Ups \approx 10$, but becomes increasingly inaccurate above this
value\cite{chen}.

\bigskip\noindent Taking both ISR and beamstrahlung effects into account, and
assuming that they take place in quick succession, the electron spectrum at the
collision point can be well approximated by a simple convolution of the two
respective spectral densities~\cite{chen,rohini}:
\bearr
f_{e|e}(x) = \int_x^1 ~\frac{d\xi}{\xi} ~f_{e|e}^{\rm ISR}(\xi) 
~f_{e|e}^{\rm beam}
(\frac{x}{\xi}) \ .
   \label{eq:convol}
\eearr
This formula can now be used to generate the spectrum illustrated in Figure~1,
which shows the (convoluted) electron luminosity for the given design parameters
at the TESLA, running at $\sqrt{s} = 800$~GeV (dashed lines) and at the CLIC,
running at $\sqrt{s} = 3.0$~TeV (solid lines). The pure ISR spectrum at TESLA
energy, which has already been remarked upon, is also shown for purposes of
comparison (dotted lines). Noting that the ordinate in this plot is logarithmic,
it is immediately apparent that beamstrahlung effects serve to broaden the energy
spectrum of the electron much more significantly than ISR alone. This energy
spread may, therefore, be used, as explained in the previous section, to excite
resonances quite some way away from the machine energy.

\bigskip\noindent To get quantitative results, we now require to substitute the
function $f_{e|e}(x)$ as obtained from Eqn.~(\ref{eq:convol}) into the general
formula in Eqn.~(\ref{eq:strucfun}) to get the final cross-section. Resonances, if
any, will then be predicted at the appropriate values of $\hat{s} = x_1x_2s$. The
remaining part of this article discusses how this can happen, and what our
expectations in respect of new physics -- specifically massive RS gravitons --
should be.

                   \section{Graviton Resonances}

\noindent The two-brane model of Randall and Sundrum (RS) has one extra dimension
compactified on a ${\bf S}^1/{\bf Z}_2$ orbifold, and two (four-dimensional)
3-branes at the orbifold fixed points.  It also assumes finely-tuned cosmological
constants on the two branes and in the intervening space or 'bulk'. On one of
these 3-branes, which is identified with the observed Universe, we have an
effective theory where the SM is augmented by a set of Kaluza-Klein excitations of
the graviton. These behave like massive spin-2 fields with masses $M_n = x_n m_0$,
where the $x_n$ are the zeroes of the Bessel function of order unity, $n$ is a
non-negative integer and $m_0$ is an unknown mass scale close to the electroweak
scale, which can be related to the radius of compactification $R_c$ and the
curvature parameter ${\cal K}$ of the orbifold ${\bf S}^1/{\bf Z}_2$, through the
expression
\bearr
m_0 = {\cal K} e^{-\pi{\cal K}R_c}
\eearr
where $R_c$ is the compactification radius of the orbifold and $e^{-\pi{\cal
K}R_c}$ is the famous exponential 'warp factor' of the RS model. We treat $m_0$ as
one of the free parameters of the model. Current experimental data from the
Drell-Yan process at the Tevatron constrain $m_0$ to be more-or-less above
160~GeV~\cite{tevatron}.

\bigskip\noindent The other undetermined parameter of the theory is the curvature
of the fifth dimension expressed as a fraction of Planck mass and is given by $c_0
= {\cal K}/M_P$. Feynman rules for the Randall-Sundrum graviton excitations can
then be read off from the well-known Feynman rules for the large extra dimension
model given in Ref.\cite{Han} by making the simple substitution $\kappa \to
4\sqrt{2}\pi c_0/m_0$. Noting that the massive graviton states exchanged in the
$s$-channel can lead to Breit-Wigner resonances, a long but straightforward
calculation must be undertaken to find the cross-section for the process $e^+ ~e^-
\to \mu^+ ~\mu^-$. We have performed this calculation and incorporated our results
in a simple Monte Carlo event generator to get numerical estimates. \\

\begin{figure}[htb]
\begin{center}
\includegraphics[height=5.2in, width = 6.2in]{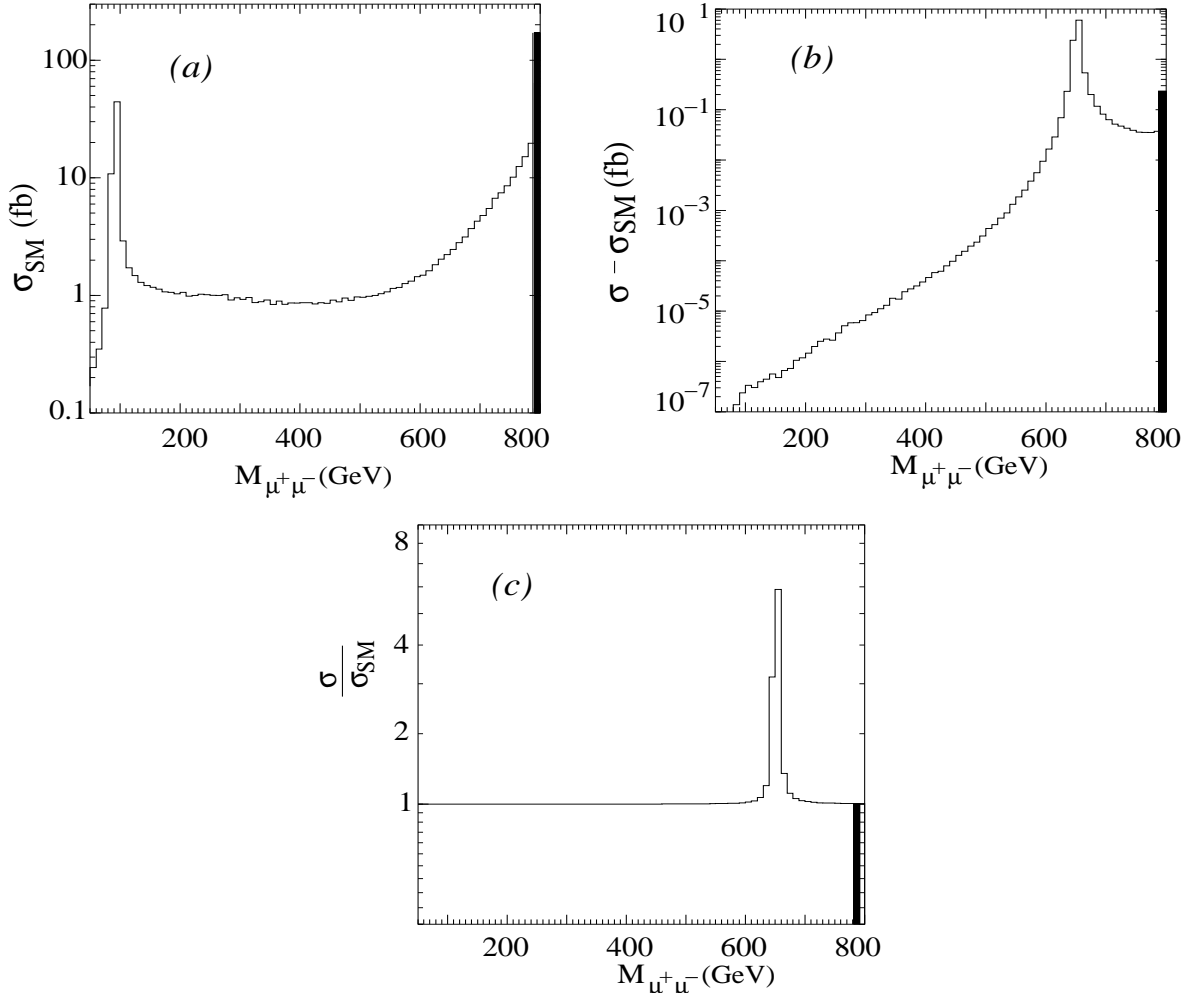}
\vskip -10pt
\caption{\footnotesize\it Invariant dimuon mass distribution with (histogram)  
and without (solid vertical bar) radiative effects. The figures correspond to {\em
(a)}~the bin-wise cross-section $\sigma_{SM}$ for the SM background,~ {\em
(b)}~the bin-wise excess cross-section $(\sigma - \sigma_{SM})$ for the RS
contribution alone and~ {\em (c)}~the signal-to-background ratio
$\sigma/\sigma_{SM}$. 'Bin-wise' cross-sections refer to integrated cross-sections
in each bin of $M_{\mu^+\mu^-}$.}
\end{center}
\end{figure}

\noindent For our numerical analysis we have chosen the parameters of the RS model
to be $m_0 = 170$~GeV and $c_0 = 0.01$, which implies that the lightest ($n = 1$)
massive excitation has mass $M_1 \simeq 652$~GeV, placing it just beyond the
present detection reach of Run-2 data at the Tevatron~\cite{tevatron}. With this
choice, however, the next excitation is predicted to have mass $M_2 > 1$~TeV,
which would put it well beyond the kinematic reach of an 800 GeV machine. At the
TESLA or even the ILC, therefore, we expect to detect one -- and only one --
resonance.  The value of $c_0$ has initially been chosen at the lower end of the
possible range, since this leads to a longer lifetime for the Kaluza-Klein state
and hence a sharper resonance in the cross-section. It also enables us to evade
the more stringent bounds from the Tevatron which arise for higher values of
$c_0$. The effects of varying $c_0$ are shown in a later section.

\bigskip\noindent We have run the event generator for the cross section subject to
the following kinematic cuts on each of the final state muons:
\begin{center}
$\bullet$ transverse momenta: $p_T^{\mu^\pm} > 20$~GeV
\hskip 20pt
$\bullet$ scattering angle: $ 10^0 < \theta_{\mu^\pm} < 170^0$ \ .
\end{center}
These are almost standard acceptance cuts and are known to eliminate most of the
backgrounds from beam-beam interactions as well as two-photon processes. Some of
our results are illustrated in Figure 2, which shows the bin-wise distribution of
invariant mass $M_{\mu^+\mu^-}$ of the (observable) final state at the TESLA
running at $\sqrt{s} = 800$~GeV.  In Figure~2($a$), we have plotted the
distribution predicted in the SM. Figure~2($b$) shows the {\it excess} over the SM
prediction expected in the Randall-Sundrum model and Figure~2($c$) shows the
signal-to-background ratio.

\bigskip\noindent At a linear collider with a fixed center-of-mass energy (i.e. no
radiation effects), all the events should be concentrated in a single invariant
mass bin at $M_{\mu^+\mu^-} = \sqrt{s}$. In Figure~2 this corresponds to the solid
vertical bar at the right edge of the $M_{\mu^+\mu^-}$ distribution. Comparison of
Figures 2($a$) and 2($b$) --- which are plotted on vastly differing scales ---
show that in this 'ideal' case, the expected signal is very small indeed, about 1
in $10^4$.  Naturally the ratio indicated by the solid bar in Figure 2($c$) is
almost precisely unity. This is not surprising, since our parameter choice leads
to a graviton of mass around 652~GeV and decay width of a few GeV, which means
that the resonance lies many decay widths away from the collision energy $\sqrt{s}
= 800$~GeV.

\bigskip\noindent The situation is not so bad, however, as the solid bars would
seem to indicate. This is, in fact, the main thrust of our work. The histograms in
Figure~2 show the invariant mass distribution {\it when radiative effects are
included}. It is immediately apparent that the effective centre-of-mass energy
$\sqrt{\hat{s}} = M_{\mu^+\mu^-}$ is spread out from the beam energy $\sqrt{s}$.
In ($a$), where only the SM prediction is plotted, we can see a distinct peak at
the lower end which showcases the `return to the $Z$-peak'. The cross-section for
this peak is not as high as has been seen at LEP and SLD in the case of a narrow
$Z$ resonance, since in the present case the peak corresponds to very small values
of the energy fraction $x$ of the electron ($x \simeq 0.11$), for which the
electron luminosity is extremely small ($f_{e|e}(x)\simeq 0.035$), as can be seen
from Fig.~1. Nevertheless, it may be seen that the peak is still higher than the
cross-section at the design energy $\sqrt{s}$.

\bigskip\noindent Since there are no other massive resonances in the SM, the shape
of the rest of the histogram in Fig.~2($a$) simply reproduces the electron
luminosity curve shown in Figure~1. In Fig.~2($b$), however, where the RS model
predictions are considered, the radiative return to a resonant 652~GeV
Kaluza-Klein mode of the graviton is quite apparent. This resonance lies much
closer to the machine energy, and hence is excited for a much larger value of
energy fraction ($x \simeq 0.82$), where the electron luminosity ($f_{e|e}(x)
\simeq 0.3$) is significantly larger. It may be pointed out that the scales along
the ordinate are quite different in Fig.~2($a$) and 2($b$), showing that the
graviton peak is roughly an order of magnitude smaller than the $Z$ peak, despite
the larger electron flux. The histogram in Fig.~2($c$), shows the
signal-to-background ratio. This ratio removes the $Z$-peak and throws the
graviton resonance into prominence, presenting us with a clear signal for a new
resonant particle.

\bigskip\noindent It is important to note that although we have concentrated on a
massive graviton resonance, the above analysis requires only the existence of a
(neutral) resonant particle, with mass in the relevant energy range and coupling
to both $e^+e^-$ and $\mu^+\mu^-$ pairs.  Thus, the same technique may be used
effectively to study
\begin{itemize}
\item extra $Z'$ bosons arising in models with extended gauge symmetries
\cite{Rizzo};
\item Kaluza-Klein excitations of $Z$-boson in models with universal
extra dimensions\cite{Kundu} ;
\item leptophilic scalars, such as dileptons and the sneutrinos of
$R$-parity violating supersymmetry with $LLE$-type operators\cite{DebSak10e}.
\end{itemize}
All of the above theories would predict similar resonances. If, indeed, such an
effect is seen, the question would at once arise as to the nature --- especially
the spin --- of the resonant (bosonic) particle. For example, only the
confirmation of two units of spin would establish the existence of a massive
graviton resonance. This question is taken up in the next section.

\bigskip\noindent Before concluding the present section, we briefly discuss an
important issue, especially from the point of view of the experimentalist, viz.
the choice of a $\mu^+\mu^-$ final state.  This is well known to be one of the
cleanest signals, but -- especially for gravitons -- this channel is suppressed by
the low branching ratio for $G_n \to \mu^+\mu^-$ compared with the branching ratio
to a pair of hadronic jets, i.e. $G_n \to q\bar q$ or $gg$.  However, in looking
for resonances in two-jet final states we must remember that two-photon processes
can give rise to a substantial dijet production rate for invariant masses quite a
bit smaller than the nominal centre-of-mass energy $\sqrt{s}$ of the
collider~\cite{dreesRG,rohini}. This would heavily contaminate the signal and may
render the results inconclusive. This argument applies for $\tau^+\tau^-$ final
states as well, since the $\tau^\pm$ decay to (narrow) jets. The $e^+e^-$ final
state will have a large contribution from the $t$-channel exchanges in Bhabha
scattering, against which background the feeble signal may be totally lost. A
similar argument holds for the $ZZ$ final state, which can be nicely reconstructed
from the leptonic decays of the $Z$, but is heavily contaminated by $Z$ pair
production from $t$-channel electron exchange diagrams. It turns out, therefore,
that the $\mu^+\mu^-$ channel chosen in the present work is indeed the best one to
trigger on in spite of the large suppression caused by the low branching ratio.

              \section{Identifying Resonant Particles}

\noindent In the previous section we have demonstrated how radiative effects like
ISR and beamstrahlung could be effectively utilised to highlight a resonance in
$e^+e^-$ annihilation. Specifically, we showed that for a RS graviton with a mass
which is accessible kinematically, one would see a distinct peak in the invariant
mass distribution of the $\mu^+\mu^-$ pair in the final state. However, merely
identifying a resonance --- even if the width can be measured and is found to fit
the theory --- does not automatically indicate whether we have seen a spin-2
particle, for it could very well be an exotic vector or scalar particle (of the
types listed above, or even of completely unknown type) of the same mass and
comparable decay width. One requires, therefore, to make some extra measurements
to identify the nature of the resonance.

\bigskip\noindent One of the important features which distinguishes a graviton
resonance from most of the vector or scalar resonances is the fact that it has,
for the same mass and couplings of comparable strength, a greater decay width.
This is essentially because the graviton couples to all particles with a coupling
strength which is determined only by the spacetime quantum numbers and is blind to
all the internal quantum numbers, i.e. the massive graviton can decay into {\it
any} pair of particles, so long as it is kinematically possible. The presence of
more channels thus enhances the decay width, even when the overall coupling is the
same as those in the case of other resonances. We may, therefore try to identify
{\it broad resonances} as a distinctive feature of RS graviton resonances.

\begin{figure}[htb]
\begin{center}
\includegraphics[width=3.6in]{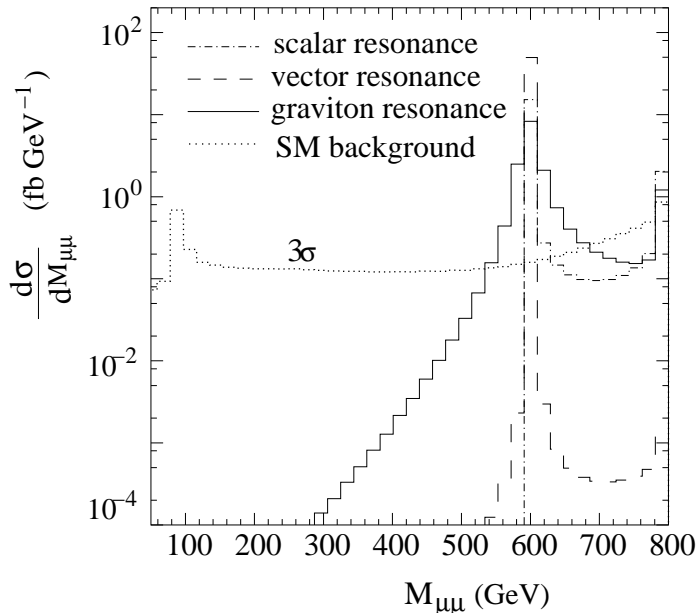}
\end{center}
\vskip -20pt
\caption{\footnotesize \it Bin-wise invariant $\mu^-\mu^-$ mass distribution for,
respectively, scalar (dash-dot line), vector (dashed line) and RS graviton (solid
line) resonances of similar mass and {\em total} cross-section. Radiation effects
are included, and the dotted histogram indicates the $3\sigma$ SM background.}
\label{fig:invM_all}
\end{figure}
\vskip -1pt
                                                                                
\bigskip\noindent To make a detailed analysis, we consider, as before, the process
$e^+ ~e^- \to X^* \to \mu^+ ~\mu^- $ where $X$ can be, respectively, a generic
spin-0 (scalar) particle or a spin-1 (vector)  particle or a RS graviton {\it of
the same mass}. We chose a fixed mass of 600 GeV for all the above particles and,
in every case, ran an event generator with the same kinematic cuts as above. The
cross-sections vary for the different exchange particles, depending on the
strength of interactions and the width of the decaying particle. We have,
therefore, taken values for the couplings which, while remaining consistent with
present experimental bounds\cite{pdb6} for the given mass, are chosen such that
the {\it total} cross-sections for the different cases (spin-0, spin-1 and spin-2)
are all equal. In Figure~\ref{fig:invM_all} we show the bin-wise distribution of
invariant mass $M_{\mu^+\mu^-}$ of the (observable) final state for the different
cases. Figure~3 is thus similar to Figure~2($b$), except that all three types of
particles are included instead of just RS gravitons. A glance at the figure shows
that there is indeed a significant difference in the decay width, with the scalar
resonance being very narrow and the graviton resonance being much broader. We
note, however, that it is the excess signal over the SM background that is of
significance and in this case, the background is quite large. In fact, if we plot
the $3\sigma$ fluctuation in the SM with 500~fb$^{-1}$ luminosity, it is seen that
only the tip of the resonance peaks would show up, and it is likely, therefore to
be very difficult to distinguish between these little bumps over the SM
background, considering that there would be large error bars in the experimental
data.

\begin{figure}[htb]
\begin{center}
\includegraphics[width=3.3in]{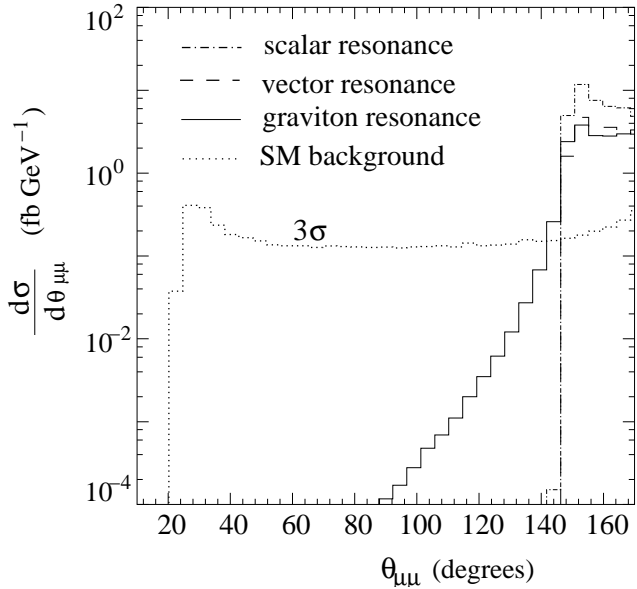}
\end{center}
\vskip -20pt
\caption{\footnotesize \it Bin-wise opening angle distribution for the final state
$\mu^+ \mu^-$ pair, for scalar, vector and RS graviton resonances of similar Mass
and total resonant cross-section. The conventions followed are the same as in
Figure~\ref{fig:invM_all}. }
\label{fig:opnA_all}
\end{figure}
\vskip -10pt

\bigskip\noindent A more direct measurement aimed at identifying the spin of the
resonant particle is to plot the angular distribution of the final state
particles, since it is well known that the spin of the decaying resonance leaves a
distinct imprint on this angular distribution. However, we must note that the
$e^+$ and $e^-$ no longer collide in the centre-of-mass frame because of radiation
effects.  It may seem, on first thoughts, that the resultant boost will not be so
significant for collisions at 800~GeV if a 652~GeV resonance is produced, but this
is not, in fact, the case. To show that the collision does, in fact, take place in
a highly-boosted frame, we have plotted the distribution for the opening angle
$\theta_{\mu\mu}$ of the final state muons in Figure~\ref{fig:opnA_all}. In the
absence of radiation effects one would expect the muons to scatter back-to-back.
i.e. $\theta_{\mu\mu} = 180^o$.  However, the energy carried away by radiated
photons makes the final state $\mu^+ \mu^-$ pair scatter at different angles so
that the {\it overall} four-momentum is conserved. Figure~4 clearly shows that
this effect is not only significant, but dominates the much weaker differences due
to spin, rendering the three histograms for, respectively, scalar (dot-dash),
vector (dashed) and RS gravitons (solid) quite indistinguishable. As in Figure~3,
the long tail of the distribution in the RS case is quite lost in the SM
background, even at the $3\sigma$ level.

\begin{figure}[htb]
\begin{center}
\includegraphics[width=3.3in]{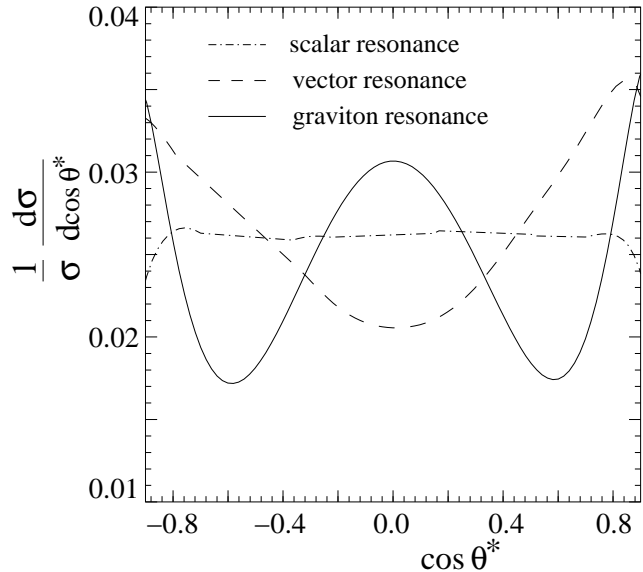}
\end{center}
\vskip -20pt
\caption{\footnotesize \it The angular distribution of the final state muon in the
rest frame of the decaying particle with respect to the boost axis, where the
boost is due to radiated photons, for scalar, vector and RS graviton resonances
respectively. The line styles are the same as in the previous two figures, but the
ordinate shows the normalised differential cross-section rather than the bin-wise
cross-section.}
\label{fig:angle_all}
\end{figure}
\vskip -10pt
                                                                                
\bigskip\noindent All is not lost, however, for we can still reconstruct the
angular distribution of the muons {\it in the rest frame of the decaying
particle}. To achieve this, we simply calculate the longitudinal component of the
total momentum of the two muons $p_L^+ + p_L^-$ and use it to compute the boost
experienced by their parent particle $X$ with respect to the laboratory frame. We
then boost back the decay products (muons) to the rest frame of $X$ and plot the
muon angular distribution in this reconstructed rest frame.
Figure~\ref{fig:angle_all} shows this (normalised) angular distribution, where
$\theta^*$ is the angle of the final state $\mu$ with respect to the boost axis.
The scalar (dot-dash), vector (dashed) and RS graviton (solid) distributions are
now quite different, with the curves in Figure~\ref{fig:angle_all} representing
the first three Legendre polynomials in $\cos \theta^*$, which is expected in view
of the appropriate angular momentum representations.  The dips in the
cross-section for a scalar resonance at the extreme values $\cos\theta^* = \pm 1$
are artifacts of the kinematic cuts mentioned in Section 3. Though we have not
shown it explicitly in the figure, we estimate that the typical error bars
expected for 500~fb$^{-1}$ luminosity should be small enough to enable the three
cases to be clearly distinguished. This reconstruction technique is, therefore,
fairly powerful and can be used to identify the spin of the resonant particle
quite unambiguously.

\bigskip\noindent One obvious spoiler for the above technique would be a situation
where there are {\it two} or more particles with the same mass but different spin.
Though this is quite unlikely in most extensions of the SM, it is far from
impossible in universal extra dimensions models\footnote{which are similar and
closely related to, but distinct from, the RS model\cite{UED}}, where the
tree-level masses of the Kaluza-Klein excitations of all SM particles are
determined by the size of the extra dimension. Fortunately, the only resonances
which can be excited in $e^+e^-$ collisions in this model are vector-like
(excitations of the photon and the $Z$), so their presence would pose no problems
for graviton identification.
                                                                                
              \section{Higher Energies and Luminosities}

\noindent Apart from the spin, the other really distinguishing feature of
Kaluza-Klein excitations is the fact that there exists a tower of such states with
successively increasing mass. The observation of such multiple spin-2 excitations
would be a clear indication of a theory with extra dimensions. In the case of the
RS model, we know that these masses vary as the zeros $x_i$ of Bessel functions of
order unity, which means that the first three resonances satisfy the ratio
\bearr
M_1:M_2:M_3 \approx 1:1.83:2.65 \ .
\eearr
If one could, for example, detect three resonances and establish that their mass
ratio satisfies this condition, we would have gone a long way toward establishing
the truth of the RS model. It, is, therefore, important to detect more of the
graviton resonances, apart from the lightest one which has been the subject of
most of the preceding discussions.

\bigskip\noindent It is practically a theorem in high energy physics that increase
in the machine energy and/or the luminosity automatically results in an increase
in the power of the machine to search for new physics. This is very much
applicable to the search for resonances. In the present case, if we consider the
RS model with $m_0 = 170$~GeV, we have already seen that the first graviton
resonance at about 652~GeV is all that is accessible at a 800~GeV machine.  At the
500~GeV option of the ILC, the possibility of exciting RS graviton resonances is
already ruled out by the Tevatron data. The next two resonances, for $m_0 =
170$~GeV, lie at around 1193~GeV and 1727~GeV respectively, and would be even
higher if $m_0$ is higher. We see, therefore, that the higher graviton resonances
are fairly heavy and can be excited only at a machine running at a significantly
higher energy.

\begin{figure}[htb]
\begin{center}
\includegraphics[width=3.5in]{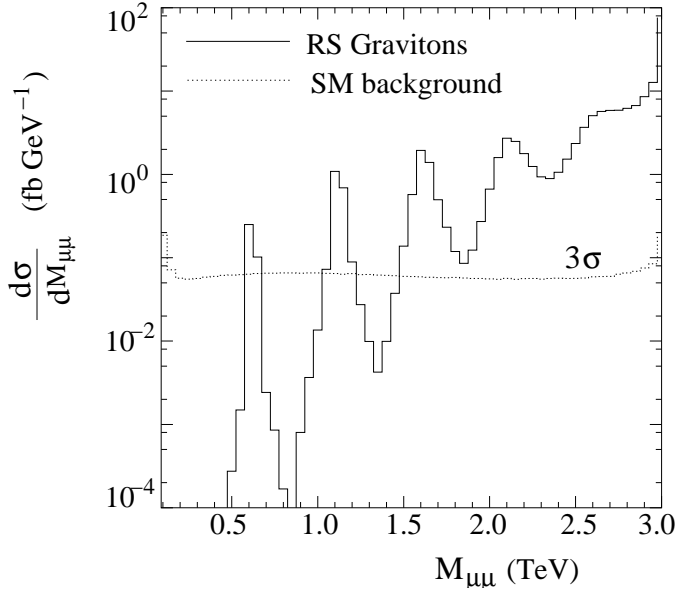}
\end{center}
\vskip -20pt
\caption{\footnotesize \it The invariant mass $M_{\mu^+\mu^-}$ distribution at the
CLIC collider running at $\sqrt{s}= 3$~TeV for RS graviton resonances (solid
line). As in Fig.~2, the RS model parameters are chosen to be $m_0 = 170$~GeV and
$c_0 = 0.01$. The dotted histogram shows, as in the preceding figures, the
$3\sigma$ SM background.}
\label{fig:clic_invM}
\end{figure}

\bigskip\noindent Of all the $e^+e^-$ machines of linear collider type proposed,
the CLIC, running at $\sqrt{s}= 3$~TeV would be the most likely one where these
higher graviton resonances may be seen. In fact, one of the positive features (so
far as this analysis is concerned) of the CLIC is that radiation effects are
considerably enhanced, the beamstrahlung parameter being as high as $\Ups = 8.1$
for the CLIC design parameters at $\sqrt{s}= 3$~TeV. This, is turn, would cause a
much larger energy spread when compared with the TESLA prediction (see
Figure~\ref{fig:bmsisr}). We make use of this fact and illustrate in
Figure~\ref{fig:clic_invM} how the multiple resonances in the RS model could be
excited at the CLIC\footnote{In our numerical analysis the kinematic cuts were
kept identical with those imposed at $\sqrt{s} = 800$~GeV, a feature which may
need reconsideration in a more realistic study.  It suffices, however, for our
purposes here, which are chiefly illustrative, to keep the same kinematic cuts.}.
Figure 6 clearly shows three resonances, with a hint of a fourth. If the
qualitative features of this prediction are reproduced by the actual data, it will
have to be followed up by two checks, namely ($a$) that the successive peaks are
in the ratio $1:1.83:2.65$, and ($b$) that the events concentrated around each of
the peaks follow the spin-2 angular distribution illustrated in Fig.~5. If {\sl
all} of these are confirmed, we can claim a `smoking gun' signal for the RS Model.
If no such effects are seen, the RS model is disfavoured but not ruled out, since
it would simply mean that the lower bound on $m_0$ moves up to around 780~GeV in
place of the present lower bound of around 160~GeV. If {\it some} of these
predictions are confirmed and others are not, it would be a challenge to the high
energy physics community to find a reasonable explanation, and it is quite
possible that in any such explanation extra dimensions will play an important
role.

\vskip -1pt
                                                                                
\bigskip\noindent It could be the case, as mentioned above, that $m_0 > 780$~GeV
and hence graviton excitations are too massive to be accessed even at the CLIC. In
fact, as the graviton mass scale $m_0$ increases, the peaks in Figure~6 shift to
the right of the graph and eventually move out of the energy reach of the machine.
It is also possible that the graviton resonance peaks become too broad to be
identified as such. It is, thus, a relevant question to ask how far the parameter
space can be probed at the design energies and luminosities. To take up this
question, we set up as a discovery paradigm the situation when the invariant mass
distribution shows a significant deviation from the SM prediction. To quantify
this, we calculate a bin-wise-summed variance $\chi^2$ in the $M_{\mu\mu}$
distribution
\be 
\chi^2(m_0,c_0)  = \sum_i
\frac{\left(N_i^{(SM+RS)} - N_i^{(SM)}\right)^2}{N_i^{(SM)}} = {\mathcal L} 
\sum_i
\frac{\left(\sigma_i^{(SM+RS)} - \sigma_i^{(SM)}\right)^2}{\sigma_i^{(SM)}} 
\ee 
where ${\mathcal L}$ is the integrated luminosity, $N_i$ denotes the number of
events in a bin and $\sigma_i$ is the cross-section in that bin. In the
denominator, we take only the Gaussian errors and neglect systematic effects
(which are completely unknown at this early stage in linear collider development).
To observe a deviation at 95\% confidence level, we require this $\chi^2(m_0,c_0)$
to be greater than a fixed number, which is, for example, 23.46 for 20 bins in
$M_{\mu\mu}$. We may, therefore, map out the entire $m_0$--$c_0$ plane for
different energies and luminosities, calculating the radiation effects in each
case using the design beam parameters given in the literature. Our results are
shown in Figure~7, where the regions to the left of and above the curves
correspond to an observable effect. For comparison, the current bounds from the
Tevatron Run-2 data are also given as a shaded region, with the unshaded region
indicated at the lower left corner being an extrapolation from the published
result. It is clear from the graph that the TESLA, running at 800~GeV, can only
improve on the Tevatron bounds marginally, that too for the high luminosity
option, increasing the discovery reach by about 150~GeV. This is quite as one
might expect, given that the mass of the lightest graviton is already known to lie
above 600~GeV. The CLIC, on the other hand, will improve the discovery reach
greatly, as the wide regions to the left of the corresponding curves indicate.

\begin{figure}[htb]
\begin{center}
\includegraphics[height=3.8in,width=4.8in]{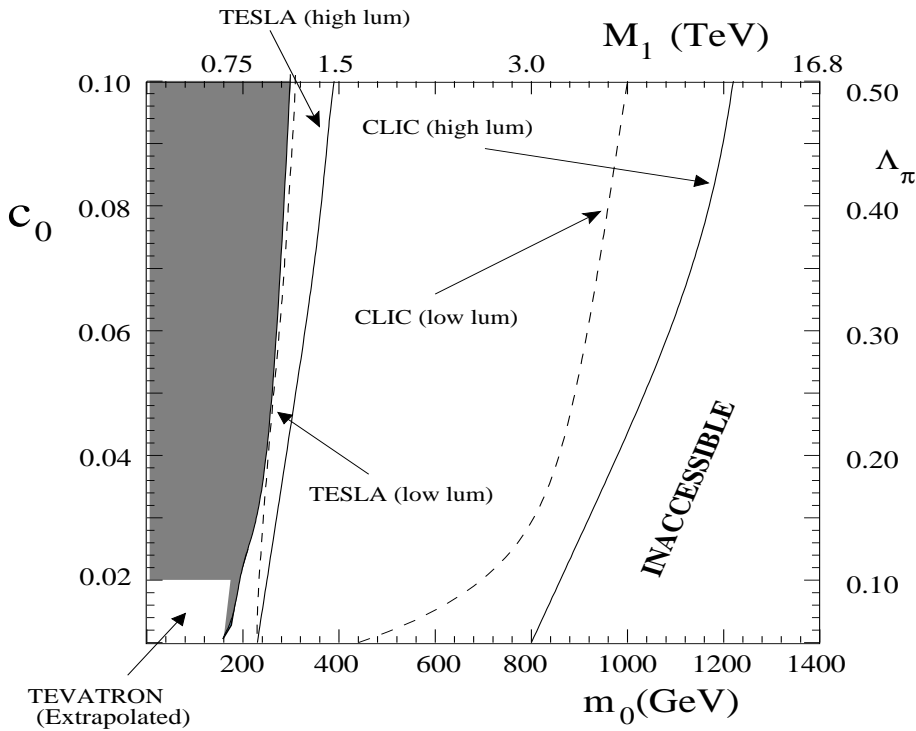}
\caption{\footnotesize\it Discovery limits in the $c_0$--$m_0$ plane for RS
gravitons at different machines, assuming different centre-of-mass energies and
integrated luminosities. Regions to the right of the curves are inaccessible to
the machine in question.  The dark shading represents the Tevatron bound from
Run-2 data, with the small unshaded region at the bottom representing an
extrapolation from the published bounds (which go down to only $c_0 \simeq 0.02$).
The dashed lines represent the regions accessible to TESLA and CLIC in the low
luminosity options, while the solid lines represent the corresponding high
luminosity options. The scales marked on the upper and right margins of the graph
represent the alternative parametrization in terms of $M_1$ and $\Lambda_\pi$
which are also found in the literature.}
\end{center}
\end{figure}
\vskip -10pt

\noindent We have deliberately avoided showing possible LHC bounds in Figure~7. At
the LHC, such bounds would arise from the combined study of a whole set of
different processes, most of which have been investigated separately in the
literature in detail. However, these analyses use different assumptions about the
luminosity and detection efficiency and even employ theoretical formulations of
some of the tricky issues in the calculations (such as the handling of
transPlanckian events).  Until a consolidated study is available in the
literature, which takes the union of all the accessible regions\footnote{like the
studies available for the supersymmetric Higgs parameter space, for example.},
therefore, it does not seem very meaningful to put LHC discovery limits in the
graph. However, a qualitative statement (perhaps more of an educated guess) is
definitely in order: the LHC discovery limits will definitely better those of the
TESLA and the ILC and may be comparable with those possible at the lower
luminosity option of the CLIC, especially for lower values of $c_0$ where the
resonances are sharp. The boundaries of the accessible region will probably be
differently shaped from the ones shown for linear colliders, with the LHC
performance being worse for large $c_0$, where the resonances are broad. More than
this, one cannot say without performing a consolidated, in-depth analysis.

                  \section{Summary and Conclusions}          

\noindent To summarise, the present study has two objectives. The first of these
is to establish -- or rather, emphasize -- the fact that at a high energy $e^+e^-$
collider, ISR and beamstrahlung can play a crucial role in the identification of
new physics effects. This is a positive feature of these radiative phenomena,
which is being now increasingly recognised as an important feature distinguishing
such machines from low-energy colliders such as the LEP and its predecessors. The
particle to be discovered can be of any kind, provided it is a neutral boson, as
demanded by the total quantum numbers of the initial $e^+e^-$ state. As our work
is of illustrative nature, we have focussed on the cleanest final state, viz.,
$\mu^+ \mu^-$, even though the other final states may also play a useful role. The
second objective of our work is to apply this analysis to the case of massive
graviton resonances. Earlier studies of such graviton resonances at $e^+e^-$
colliders have assumed that the linear collider will run over a
continuously-varying range of energies, which is almost certainly not going to be
the case. We find that we are able to recover all the features of those earlier
analyses simply by including radiation effects at a fixed machine energy. Not only
do we find the resonant `bumps', but we can also measure the spin of the resonant
particle, even in the presence of large missing energy. This is achieved by the
simple technique of reconstructing the `boson' rest frame, a practice borrowed
from the hadron collider toolkit.

\bigskip\noindent If we consider the actual results obtained for the RS graviton
search, it must be admitted that the results, except for those predicted at the
CLIC, though useful enough, are not terrifically exciting. Since RS gravitons are
already known to be heavier than 650~GeV, we have performed most of our analysis
at the 800~GeV option of the TESLA, where the beam parameters are clearly
enunciated, rather than at the 1~TeV option of the ILC, which is still at the
planning stage. The actual kinematic range which can be probed is, therefore,
already rather small, lying somewhere between 650~GeV and about 780~GeV. This
cannot be helped, since the TESLA/ILC designs are intended for a different
purpose, viz. precision measurements of physics that will be discovered
(hopefully) at the LHC. At the CLIC, however -- assuming that it will be built
some day -- we find that graviton searches take on an entirely different level of
importance. In fact, for graviton searches, CLIC would far surpass machines like
the Tevatron, the TESLA/ILC, and may do significantly better than the LHC. Our
results may, therefore, be considered a strong motivation to build a multi-TeV
linear collider at some time in the not-so-remote future. Till then, we must be
satisfied with the superior ability of linear colliders of lesser energy to make
precision measurements of such attributes as the `boson' spin. What is more to the
point is that if graviton resonance(s) are lying just around the corner, as it
were, then the LHC is likely to detect them, but identification of the resonances
as gravitons may prove quite difficult. In such cases, it is a linear collider
which we must turn to for the measurement of spin, and here graviton production
will be almost completely dependent on radiation effects, as shown in the present
work.

\bigskip\noindent {\sl Acknowledgments}: {\small This work was partially supported
by the Department of Science and Technology, India, under project number
SP/S2/K-01/2000-II. The authors would also like to thank the organisers of the
{\it Eighth Workshop on High Energy Physics Phenomenology (WHEPP-8)} at Mumbai,
India (January 2004), where the idea for making the present study originated.}


\end{document}